\documentclass[%
reprint,
superscriptaddress,
amsmath,amssymb,
aps,
pra,
]{revtex4-2}
\usepackage{graphicx}

\usepackage{dcolumn}
\usepackage{bm}
\usepackage{bbold}
\usepackage[export]{adjustbox}
\usepackage{float}

\usepackage{mathtools} 

\DeclarePairedDelimiterX\braket[2]{\langle}{\rangle}{#1 \delimsize\vert #2}

\usepackage{array,colortbl,xcolor}

\usepackage[utf8]{inputenc}
\usepackage[T1]{fontenc}
\usepackage{mathrsfs}
\usepackage{dsfont}

\usepackage{amssymb}
\usepackage{amsmath}
\usepackage{amsfonts}
\usepackage{amsthm}

\usepackage{hyperref}
\hypersetup{pdfpagemode=UseNone}

\newcounter{rem}
\newcommand{\mc}[1]{\mathcal{#1}}

\def\>{\rangle}
\def\<{\langle}
\newcommand{\proj}[1]{| #1 \rangle\! \langle #1 |}
\renewcommand{\rho}{\varrho}
\newcommand{\idty}{\mathds{1}}
\def\tr{{\rm tr}}

\def\textbf#1{{\bf #1}}

\newcommand{\Nl}{\mathbb{N}}

\usepackage{xcolor}

\usepackage{soul}

\begin{document}

\title{ Approximating Invertible Maps by Recovery Channels:\\ Optimality and an Application to Non-Markovian Dynamics}
\author{\ Lea Lautenbacher}
\email[]{lea.lautenbacher@uni-ulm.de}
\affiliation{Departamento de F\'{\i}sica, Universidade Federal de Pernambuco,
Recife, PE  50670-901 Brazil}
\affiliation{Institut f\"ur Theoretische Physik, Albert-Einstein-Allee 11, Universit\"at Ulm, D-89069 Ulm, Germany}
\author{\ Fernando de Melo}
\email[]{fmelo@cbpf.br}
\affiliation{Centro Brasileiro de Pesquisas F\'{\i}sicas, Rua Dr.~Xavier Sigaud 150, 22290-180 Rio de Janeiro, Brazil}
\author{Nadja K. Bernardes}
\email[]{nadja.bernardes@ufpe.br}
\affiliation{Departamento de F\'{\i}sica, Universidade Federal de Pernambuco,
Recife, PE  50670-901 Brazil}


\begin{abstract}
We investigate the problem of reversing quantum dynamics, specifically via optimal Petz recovery maps. We focus on typical decoherence channels, such as dephasing, depolarizing and amplitude damping. We illustrate how well a physically implementable recovery map simulates an inverse evolution. We extend this idea to explore the use of recovery maps as an approximation of inverse maps, and apply it in the context of non-Markovian dynamics. We show how this strategy attenuates non-Markovian effects, such as the backflow of information.
\end{abstract}

\maketitle

\section{Introduction}
\label{intro}

With the advance of quantum computation and quantum communication, the interest in open quantum systems was renewed in the last years. In real-world situations no physical system is completely isolated, and this unavoidable interaction between system and its environment is responsible for damaging important quantum resources, such as coherence and entanglement. In order to minimize these detrimental effects, one can explore memory effects and backflow of information that may be present in non-Markovian dynamics \cite{Rivas_2014, Ines}. Those effects have been extensively controlled and manipulated by recent experimental techniques \cite{Zanardi,Chrus_2019,Li_2019,Li_2020}.

Another possibility to minimize decoherence effects is to endeavour the recovery of the original quantum state or even to revert the noisy process. However, in principle, only unitary dynamics can be perfectly reverted. For noisy situations, different recovery maps have been proposed \cite{Renner,Wilde1}. Especially in the context of quantum error correction, Petz recovery maps \cite{Petz1,Petz2} have been very useful to develop recovery operations \cite{Wilde2,Wilde3,Wilde4, Sutter, Barnum, Datta,Kwon,Alvaro,Mandayam}.

The evolution of a quantum system from an initial time $0$ to a later time $t$ is described by a family of completely positive and trace preserving (CPTP) maps, here denoted by $\Lambda_{t,0}$ with $t \geq 0$. As mentioned before, the inverse map, here represented by $\Lambda_{t,0}^{-1}$ in general, is not a valid physical process. There are indeed maps where the inverse is not even  mathematically well defined, known as non-invertible maps. Recently it has been shown that the non-invertibility of a map can be explored as a witness of non-Markovianity \cite{Chrus2,Dugic, Katarzyna}. Note that many interesting physical maps are non-invertible, like  a completely depolarizing channel, a completely dephasing channel, and a spontaneous emission. 

In this work, we study optimal recovery strategies exploring the Petz maps. We analyze the behavior of paradigmatic one-qubit quantum channels: dephasing, depolarizing and amplitude damping. We show how the recovery map can be easily computed and optimized. To measure how well the Petz recovery map recovers a random quantum state, we use the fidelity function as figure of merit. Our numerical analysis elucidates what is the optimum strategy in these cases, even without any initial assumption -- as the one required for instance in Ref.~\cite{Barnum}. Our analysis allows us to identify that some non-invertible maps can, in principle, be better recovered than others. 

As an application, we show how we can explore recovery maps in the context of non-Markovian maps. Markovian evolutions can be defined by family of maps that are divisible into CP maps, i.e. $\Lambda_{t,0}=\Lambda_{t,s}\Lambda_{s,0}$ for all $t\geq s \geq 0$. If the inverse of the map is well defined, the intermediate map is given by $\Lambda_{t,s}=\Lambda_{t,0}\Lambda^{-1}_{s,0}$. We explore what would be the consequences of replacing the inverse of the map (a non-physical map) by a recovery map (a physical map). We show that this new evolution still presents backflow of information (a characteristic of non-Markovian evolutions), but in a attenuated way. With this approach we elucidate a subtle characteristic of non-Markovian evolutions: divisibility of the evolution in CP maps is not enough to transform the evolution from non-Markovian to Markovian. The intermediate map $\Lambda_{t,s}$ can only depend on times $t$ and $s$, and cannot retain any information about previous times.

This paper is structured as follows: we begin by defining in Sec. II reversible, invertible and recoverable maps. In Sec. III, we optimize the Petz recovery maps for paradigmatic one-qubit decoherence channels. Sec. IV presents an application of the formalism developed in the previous sections in the context of non-Markovian dynamics. We conclude in Sec. V.  

\section{Reversible, Invertible, and Recoverable maps}
\label{maps}

The dynamics of quantum systems is generally described by a family of linear completely positive and trace preserving  maps. Such CPTP maps model, for instance, noisy dynamics (open quantum system scenario) and quantum communication channels \cite{Breuer}.

Let to a quantum system be associated a Hilbert space $\mc{H}$. The set of all possible system states is then $\mc{D}(\mc{H}) = \{\rho \in \mc{L}(\mc{H}) \;|\; \rho \ge 0,\; \tr(\rho)=1\}$,  where $\mc{L}(\mc{H})$ represents the linear operators acting in $\mc{H}$. A CPTP map $\Lambda:\mc{L}(\mc{H})\mapsto \mc{L}(\mc{H})$, also called a quantum channel,  can be characterized by a set of operators $\{K_i\}$, with each $K_i:\mc{H}\mapsto \mc{H}$ known as a Kraus operator, as follows:
\begin{equation}
    \label{eq:cp}
    \Lambda(\omega) = \sum_i K_i \omega K_i^\dagger,
\end{equation}
for all $\omega \in \mc{L}(\mc{H})$. Such a characterization guarantees the complete positivity of $\Lambda$. To be a trace preserving map the Kraus operators must abide by $\sum_i K_i^\dagger K_i = \idty$.

When the set of Kraus operators of a given CPTP map is composed by a single element, the trace preservation condition ensures the map to be unitary. In this case the mapping is reversible. Given a CPTP map $\Lambda:\mc{L}(\mc{H})\mapsto \mc{L}(\mc{H})$, we say it is \textit{reversible} if there exists another CPTP map $\Lambda^*:\mc{L}(\mc{H})\mapsto \mc{L}(\mc{H})$ such that
\begin{equation}
    \label{def:reversible}
    \Lambda^*(\Lambda(\rho)) = \rho, \; \forall \rho\in \mc{D}(\mc{H}).
\end{equation}
If $\Lambda$ is a unitary map, with Kraus operator $U$ acting on $\mc{H}$, then $\Lambda^*$ has Kraus operator $U^\dagger$. In fact, it is easy to show that a CPTP map is reversible if, and only if, it is a unitary map~\cite{Preskillnotes}. 

One way to relax the above definition is to no longer demand the map acting on the output of $\Lambda$ to be CPTP. Given a CPTP map $\Lambda:\mc{L}(\mc{H})\mapsto \mc{L}(\mc{H})$, we say it is \textit{invertible} if there exists another linear map, not necessarily CPTP, $\Lambda^{-1}:\mc{L}(\mc{H})\mapsto \mc{L}(\mc{H})$ such that
 \begin{equation}
    \label{def:invertible}
    \Lambda^{-1}(\Lambda(\rho)) = \rho, \; \forall \rho\in \mc{D}(\mc{H}).
\end{equation}
Clearly any unitary map is also invertible \cite{BookRivas}. However, maps describing noisy dynamics, as the non-fully depolarizing channel (see section~\ref{numerics}), are invertible but not reversible. This kind of maps are most commonly encountered when describing  open quantum system dynamics. Note that the fully depolarizing map is not invertible, as it sends all the input states to the maximally mixed state.

Another possible way to relax the definition of reversible maps, is by requiring it to hold only for a subset of $\mc{D}(\mc{H})$. Given a CPTP map $\Lambda:\mc{L}(\mc{H})\mapsto \mc{L}(\mc{H})$, we say it is \textit{recoverable } if there exists another CPTP map $\tilde{\Lambda}:\mc{L}(\mc{H})\mapsto \mc{L}(\mc{H})$ and  $\mc{S}(\mc{H})\subseteq \mc{D}(\mc{H})$ such that
\begin{equation}
    \label{def:recoverable}
    \tilde{\Lambda}(\Lambda(\rho)) = \rho, \; \forall \rho\in \mc{S}(\mc{H}).
\end{equation}
The map $\tilde{\Lambda}$ is dubbed the recovery map of $\Lambda$ for the subset $\mc{S}(\mc{H})$ \cite{Renner}. The unitary map is recoverable, but  there are recoverable maps that are not unitary. Other than that, sufficient conditions for the existence of recovery maps for a subset $\mc{S}(\mc{H})$ is a much studied topic~\cite{Petz1, Barnum, Renner, Wilde1, Sutter, Datta,Wilde2,Kwon}. 

The most well-known class of recovery maps are the so-called Petz recovery maps~\cite{Petz1}.  Given a CPTP map $\Lambda:\mc{L}(\mc{H})\mapsto \mc{L}(\mc{H})$, the corresponding Petz recovery map is a CPTP map defined as: 
\begin{equation}
\label{Petz}
\Lambda_{P}^\sigma(\rho) := \sigma^{\frac{1}{2}} \Lambda^\dagger\left(\Lambda(\sigma)^
{-\frac{1}{2}}\;\rho\;\Lambda(\sigma)^{-\frac{1}{2}}\right)\sigma^{\frac{1}{2}},
\end{equation}
where $\Lambda^\dagger:\mc{L}(\mc{H})\mapsto \mc{L}(\mc{H})$ is the trace-dual of $\Lambda$, defined in terms of the Kraus operators $\Lambda^\dagger(\rho) = \sum_{i}^{n}K_{i}^{\dagger}\rho K_{i}$ and $\sigma\in  \mc{L}(\mc{H})$ is a reference state. If the relative entropy does not change by the action of the map, i.e, 
\[ S(\rho||\eta)=S(\Lambda(\rho)||\Lambda(\eta))\;\; \forall \rho, \eta \in \mc{S}(\mc{H}) \]
with $S(\rho||\eta)= \tr(\rho \log \rho -\rho \log \eta)$, then there exists $\sigma$ such that $\Lambda_{P}^\sigma( \Lambda(\rho))=\rho$ for all $\rho \in \mc{S}(\mc{H})$.

The actual construction of the recovery channel, i.e., the choice of optimal reference state $\sigma$ for a given channel $\Lambda$ and set of states $\mc{S}(\mc{H})$, is only known for few ``abstract'' cases. For instance, it has been shown in Ref.~\cite{Barnum} that for an ensemble of commuting density matrices,  $\{p_i, \rho_i\}$, the optimal recovery channel is obtained with the reference state being $\rho= \sum_i p_i \rho_i$.

The aim of the present contribution is to investigate the choice of optimal Petz recovery map in physically motivated scenarios.  More concretely, given a noisy process described by a  non-CP invertible map $\Lambda$, what is the best choice of reference state $\sigma$ that makes $\Lambda_P^\sigma$ to act as close as possible to a reverse map $\Lambda^*$?

\begin{figure}[ht]
\centering
\includegraphics[width=0.4\textwidth]{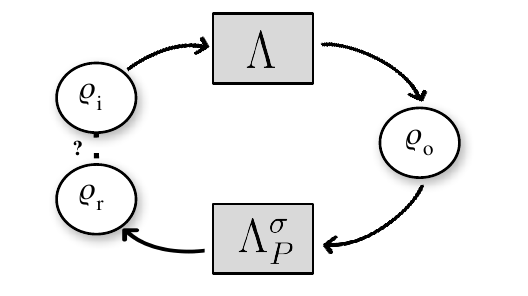}
\caption{Schematic representation of a recovery process. An initial state $\rho_{i}$ evolves through a noisy channel $\Lambda$ to a state $\rho_{o} = \Lambda(\rho_{i})$. After the evolution a Petz recovery map is applied resulting in $\rho_{r}= \Lambda_{P}^{\sigma}(\rho_{o})$ with the aim to be as close as possible from $\rho_i$.}
\label{fig:scheme}
\end{figure}

This question is schematically explained in Fig.~\ref{fig:scheme}. In the forward direction, an input state $\rho_{i}$ undergoes the action of a noisy channel $\Lambda$, leading to  the output state $\rho_{o} = \Lambda(\rho_i)$. In the backward direction, aiming at reversing the effect of the noise, we apply the channel $\Lambda_P^\sigma$ recovering the state $\rho_r = \Lambda_P^\sigma (\rho_o) $. The question is then, how close $\rho_r$ is from $\rho_i$. To quantify this closeness we use the fidelity between the states, $F(\rho_i, \rho_r) = ||\sqrt{\rho_i}\sqrt{\rho_r}||_{1}^{2}$, with the trace norm defined as $||A||_1= \tr\sqrt{A^\dagger A}$. The optimal reference state, $\sigma^*$, is then the one that maximizes the average fidelity over all input states, i.e., $\sigma^* = \text{argmax}F_\Lambda(\sigma)$, where
\begin{equation}
\label{eq:meanfid}
    F_\Lambda(\sigma) := \int d\mu_\rho \; F\big(\rho,\Lambda_P^\sigma( \Lambda(\rho))\big),
\end{equation}
with $d\mu_\rho$ a uniform measure over the input states.

\section{Optimal Petz recovery maps for paradigmatic one-qubit channels}
\label{numerics}

In this section we numerically obtain the best recovery channel for paradigmatic 1-qubit noisy channels, namely:  dephasing,  depolarizing, and  amplitude damping  channels. As we do not impose any restriction on the input states, we optimize $F_\Lambda$, Eq.\eqref{eq:meanfid}, over a uniform distribution of mixed input states. In what follows, we generate random 1-qubit mixed states, by first generating Haar random 2-qubits pure states, followed by the partial trace of the second qubit. Given that we are taking non-unitary processes, and the full state space as input, perfect recovery is impossible. Nevertheless, we can find the best Petz recovery strategy using the optimal mean fidelity as a measure of reversibility for quantum channels.

\subsection{Unital channels: Dephasing and depolarizing}
\label{sub:unital}

Unital quantum channels are those that map the maximally mixed state onto itself, $\Lambda(\idty/2)=\idty/2$. In other words,  the maximally mixed state is a fixed point for these maps. Two of the most important unital channels are the dephasing and the depolarizing noisy channels.

The dephasing channel, $\Lambda_{Deph}$, destroys the relative phase information between the computational basis states, being one the most prevalent types of noises in physical realizations. The depolarizing channel, $\Lambda_{Depo}$, can be seen as noise that probabilistically changes the system state by the maximally mixed state. Mathematically, these channels are modeled as follows:
\begin{equation}
    \label{ch:deph}
    \Lambda_{Deph}(\rho) = \left(1-\frac{p}{2}\right)\rho + \frac{p}{2}Z \rho Z;
\end{equation}
\begin{equation}
    \label{ch:depo}
    \Lambda_{Depo}(\rho) = \left(1-\frac{3p}{4}\right)\rho + \frac{p}{4}\left(X \rho X + Y \rho Y + Z \rho Z\right).
\end{equation}
In the equations above, $p\in[0,1]$ is a parameter that characterizes the noise strength ($p=0$, no noise; $p=1$ maximal noise), and $X$, $Y$, and $Z$ are the usual Pauli matrices.

\begin{figure*}[h t]
    \centering
    \includegraphics[width=.48\linewidth]{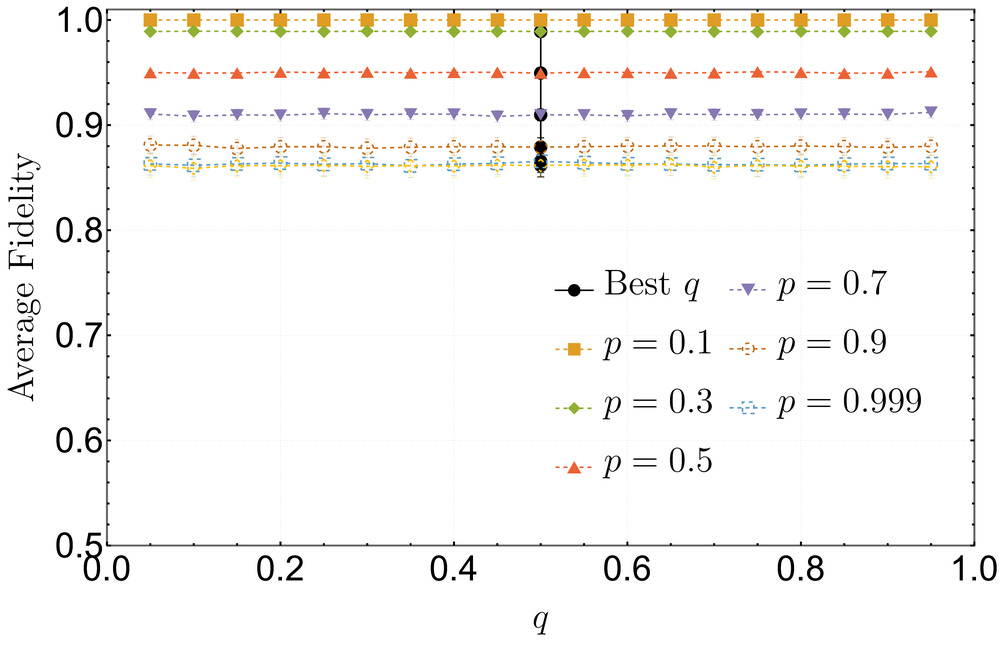}\hspace{0.5cm}
    \includegraphics[width=.48\linewidth]{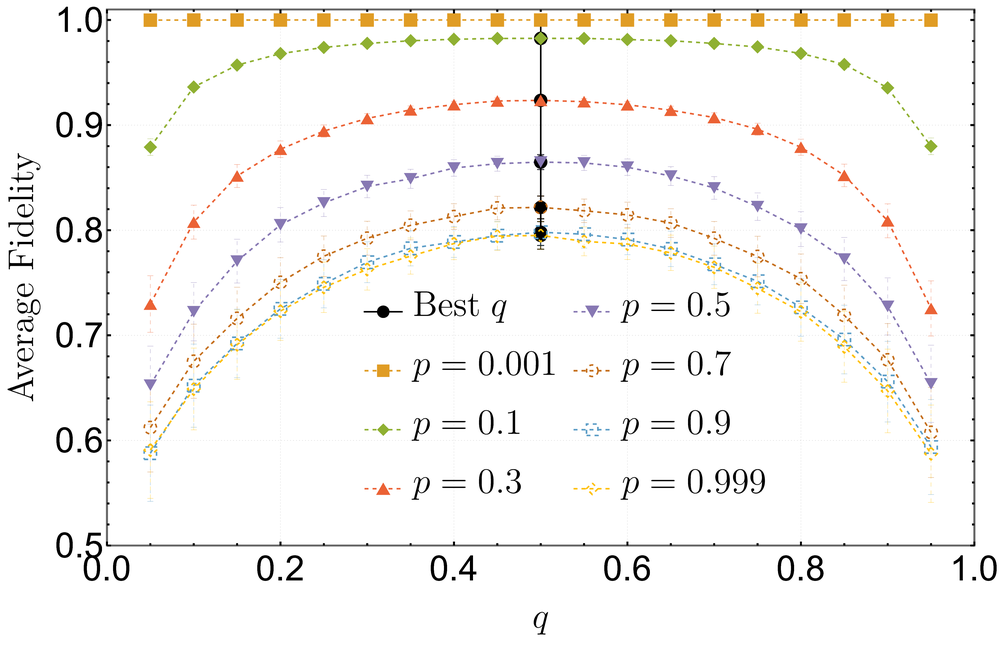}
    \caption{Average fidelity between an initial state and the recovered state via the Petz map. We took a sample of $10^4$ uniformly distributed initial mixed states. The error bars indicate the variance of the fidelity distribution. Different noise parameters $p$ were analyzed for different reference states ($\sigma=(1-q) \proj{0} + q \proj{1}$) for the dephasing map (left) and the depolarizing map (right). The optimal strategy for each noise strength is marked with solid black circle.}
    \label{fig:opt_unital}
\end{figure*}

We want to determine the optimal recovery channel for such  noise models, fixed a noise strength $p$, taking as input a uniform distribution over all mixed states. Given that we are taking a uniform measure over mixed states, and that the channels are unital, it is expected that the best Petz recovery map is obtained by taking the reference state as the maximally mixed one. This is numerically  confirmed by parametrizing the reference state as 
\begin{equation}
\label{eq:sigma}
\sigma=(1-q) \proj{0} + q \proj{1},
\end{equation}
with $q\in [0,1]$, and evaluating the average fidelity as a function of $q$ for both channels. The above parametrization is immediately suggested by the azimuthal  symmetry of the output distribution of states for both noisy channels. As it is clear from Fig.~\ref{fig:opt_unital}, taking $q=1/2$, i.e., choosing the reference state as the maximally mixed one, is always an optimal choice for the dephasing and depolarizing channels. For the dephasing channel, as all the states of the form of $\sigma$ are preserved, any value of $q$ is also an optimal choice, check Appendix \ref{ap:analytical} for the explicit calculation.

\begin{figure*}[t]
    \centering
    \includegraphics[width=.48\linewidth]{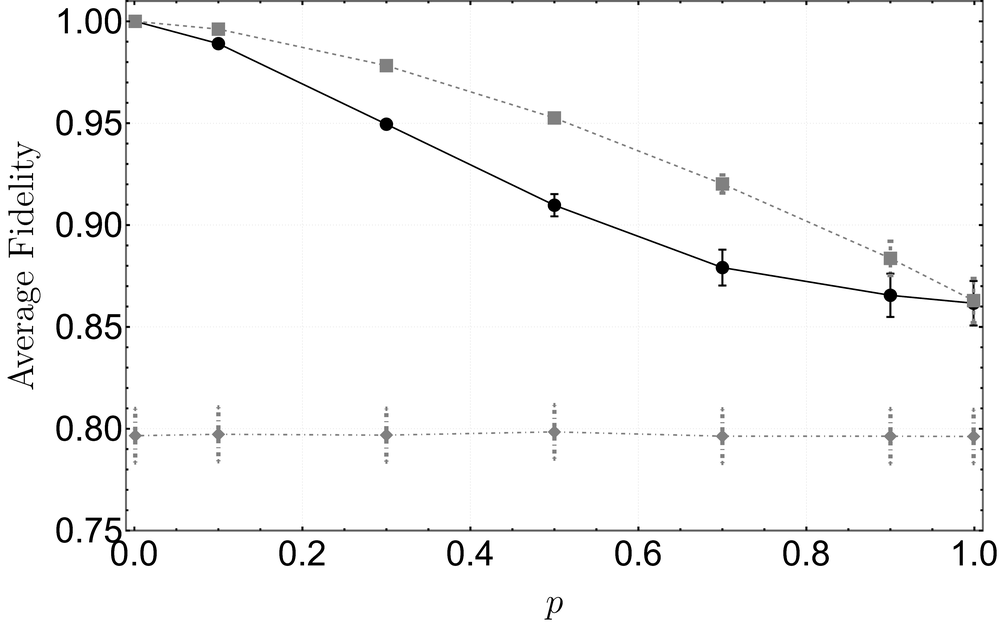}\hspace{0.5cm}
    \includegraphics[width=.48\linewidth]{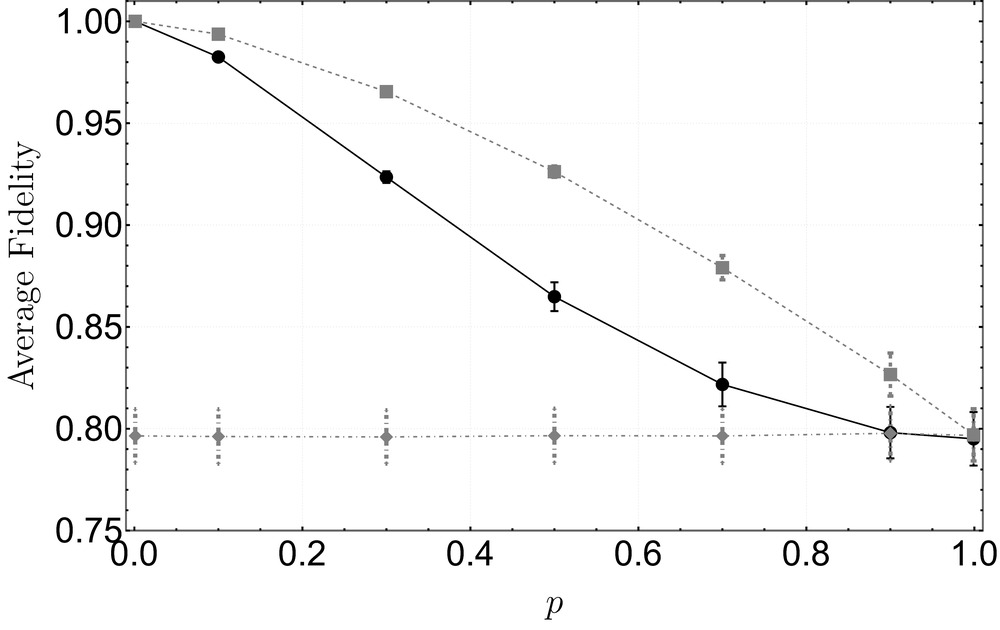}
    \caption{Average fidelity -- dephasing (left), and depolarizing (right) -- between the initial state and the state recovered by different strategies:  the identity channel (dotted line), optimal Petz map (solid line), and always returning the maximally mixed state (dot-dashed line). Plot in terms of the noise parameter $p$. We took a sample of $10^4$ uniformly distributed mixed states. The variance of the fidelity distribution is indicated by the error bars.}
    \label{fig:best_strat_unital}
\end{figure*}

Using the maximally mixed state as reference state, and the unitality of such maps, we obtain that 
\[\Lambda_P^{\idty/2}(\rho)= \Lambda^\dagger(\rho).\]
Noticing furthermore that the dephasing and depolarizing noise models are self-dual ($\Lambda^\dagger=\Lambda)$, we reach the conclusion that the optimal Petz recovery map for these noise models is again to apply the noisy map. As such, no recovery is actually obtained. In fact, it is clear from this discussion, that to take the identity channel as a ``recovery map'' is more advantageous than taking the optimal Petz when considering the set of all single qubit states as input. This conclusion is illustrated in Fig.~\ref{fig:best_strat_unital}, where we compare the optimal Petz map, for a given $p$, with two other strategies: first, the strategy of using an identity map as recovery map, i.e., to simply ``return'' the output state $\Lambda_{Deph/Depo}(\rho)$ as the best approximation for $\rho$. Second, we take as a recovery map the fully-depolarizing map, i.e., independently of the output state $\Lambda_{Deph/Depo}(\rho)$ we return the maximally mixed state. From Fig.~\ref{fig:best_strat_unital}, the latter is clearly the worst strategy, despite the fact that the maximally mixed state is indeed the average state for the input distribution of states~\cite{Barnum}. 

\subsection{Non-unital channel: Amplitude damping}
\label{sub:non-unital}

For non-unital channels, the choice of the optimal reference state for the Petz recovery map is not so obvious. As an example of non-unital channel we analyze the amplitude damping channel. Such a channel models the spontaneous decay of a two-level atom when interacting with a zero temperature environment. Mathematically, the amplitude damping channel can be written in the Kraus form as
\[
\Lambda_{AD}(\rho) = K_0(p) \rho K_0^\dagger(p) + K_1(p) \rho K_1^\dagger(p),
\]
with Kraus operators
\begin{equation}
\label{eq:Kraus_AD}
K_{0}(p) = \begin{pmatrix}
1 & 0 \\
0 & \sqrt{1-p}
\end{pmatrix},\;
K_{1}(p) = \begin{pmatrix}
0 & \sqrt{p} \\
0 & 0
\end{pmatrix}.
\end{equation}
Again the parameter $p\in[0,1]$ quantifies the noise strength. Taking $p=0$ there is no decay, while for $p=1$ the system is left in the ``ground state'' $\proj{0}$.

\begin{figure}[t]
    \centering
    \includegraphics[width=\linewidth]{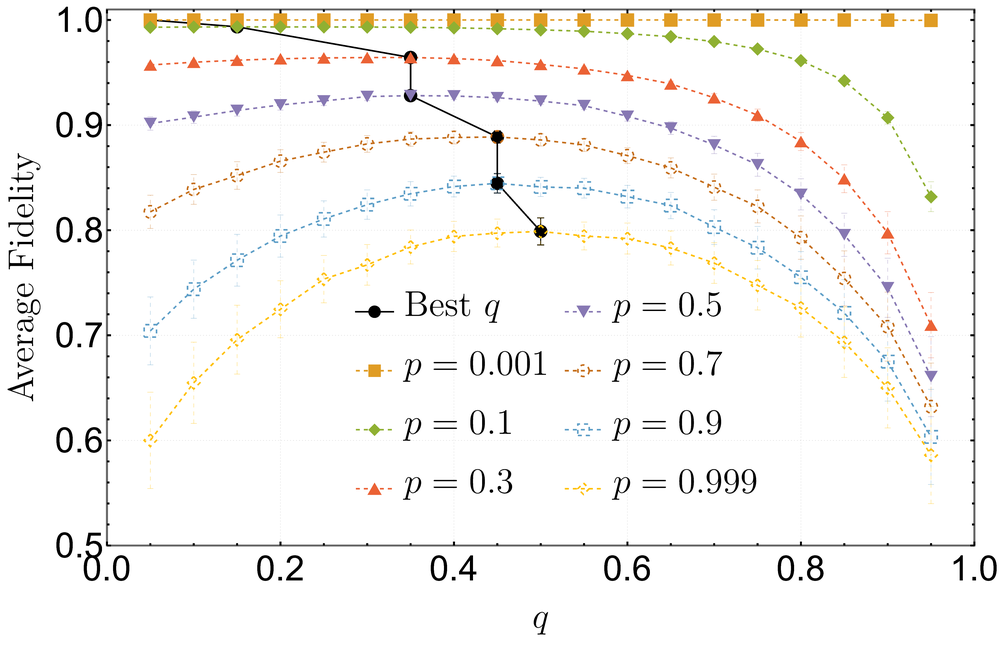}
    \caption{Average fidelity between the initial state and the recovered state via the Petz map for the amplitude damping map. We took a sample of $10^4$ uniformly distributed mixed states. The variance of the fidelity distribution is indicated by the error bars. Different noise parameters $p$ were analyzed for different reference states ($\sigma=(1-q) \proj{0} + q \proj{1}$). The optimal strategy for each noise strength is marked with solid black circle.}
    \label{fig:opt_ad}
\end{figure}

Despite the fact that the amplitude damping channel is not unital, it maps a uniform distribution of states onto a distribution which is invariant under rotations around the $z$-axis of the Bloch sphere. Given that, it is reasonable to suppose that the optimal Petz recovery channel is obtained with a reference state that is located on the $z$-axis, i.e., the optimal reference state can be written as in the state in Eq.~\eqref{eq:sigma}. In Fig.~\ref{fig:opt_ad} we use this parametrization to numerically obtain the best reference state for different values of the noise strength $p$.

\begin{figure}[t]
    \centering
    \includegraphics[width=0.9\linewidth]{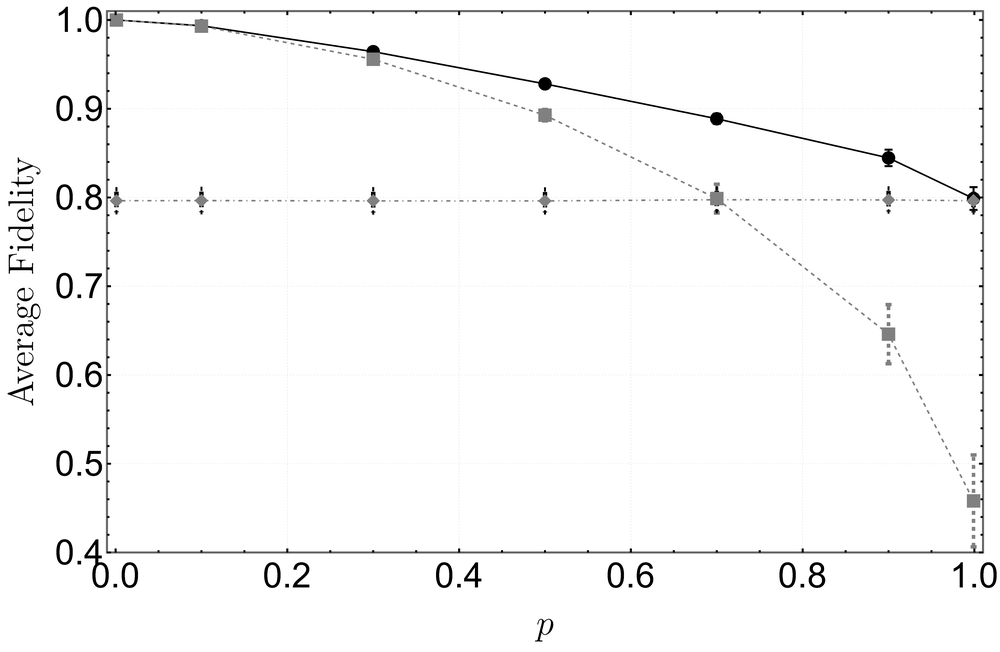}
    \caption{Average fidelity for the amplitude damping map between the initial and the state recovered by different strategies: the identity channel (dotted line), optimal Petz map (solid line), always returning the maximally mixed state (dot-dashed line). Plot in terms of the noise parameter $p$. We took a sample of $10^4$ uniformly distributed mixed states. The variance of the fidelity distribution is indicated by the error bars.}
    \label{fig:best_strat_AD}
\end{figure}

From Fig.~\ref{fig:opt_ad} we see that for weak amplitude damping, the best reference state is the $\proj{0}$. As the noise strength increases, the best reference state moves towards the maximally mixed state.  To understand that, we compare the optimal Petz map, for a given $p$, with the previous strategies: the identity map as the recovery map (always returning $\rho_r=\Lambda_{AD}(\rho)$), and returning $\rho_r=\idty/2$ independently of the initial state. The comparison among these strategies is shown in Fig.~\ref{fig:best_strat_AD}.

As it can be apprehend from Fig.~\ref{fig:best_strat_AD}, for very weak amplitude damping the best Petz recovery map (with reference state around the $\proj{0}$) is equivalent to the identity map. This is expected, as the state $\Lambda_{AD}(\rho)$ should be very close  to $\rho$ for small $p$'s. For strong amplitude damping the state $\Lambda_{AD}(\rho)$  lost almost all the information about $\rho$. In this case the optimal Petz recovery map (with reference state around the $\idty/2$) is equivalent to a fully-depolarizing map -- when no information is available, the unbiased choice is to return the maximally mixed state. It is however interesting to notice that for intermediate noise strengths the optimal Petz recovery map is better than the other strategies. 

Note that, fixing a strategy, by comparing Fig.~\ref{fig:best_strat_unital} and Fig.~\ref{fig:best_strat_AD}, the average fidelity for the dephasing map is always higher. Thus, in this context, we can conclude that the dephasing channel can be better recovered. This is related to the image set of each evolution. Because of the symmetry of the dephasing process, it can be said that some information is still available after the decoherence process.

\section{Recovery channels applied to Non-Markovian dynamics }
\label{applications}

With the results of the previous section in hands, now we exploit the impact of recovery channels in non-Markovian dynamics.

Formally, we will consider the definition of Markovianity as the divisibility of the maps. A family of  dynamical maps $\Lambda_{t, 0}$ is divisible if it can be expressed as a composition of linear maps
\begin{equation}
\label{total}
    \Lambda_{t,0} = \Lambda_{t,s} \Lambda_{s,0} \quad \text{for all}\,\, t\ge s \ge 0.
\end{equation}
The dynamics is called Markovian, or CP-divisible, if the intermediate map $\Lambda_{t,s}$ is CPTP for all $t\ge s \ge 0$ \cite{RHP}. Note that if $\Lambda_{s,0}$ is invertible, then one can obtain $\Lambda_{t,s}$ as
\begin{equation}
\label{intermediate}
    \Lambda_{t,s} = \Lambda_{t,0}\Lambda^{-1}_{s,0}.
\end{equation}
Two points should, however, be noticed. First, as the inverse of a map is in general not-completely positive (NCP), then the complete positivity of  $\Lambda_{t,s}$ is not guaranteed. Second, for Eq.~\eqref{intermediate} to be consistent, $\Lambda^{-1}_{s,0}\Lambda_{s,0} = \mathbb{1}$, as defined in Eq.\ref{def:invertible}.

Usually non-Markovian dynamics are associated with a backflow of information~\cite{Breuer2,RHP}. One of the common witnesses of this backflow of information is the distinguishability between two states. Thus, a subset of states will become relevant and the use of a Petz recovery map is then justified in this scenario.

Here we will then employ Petz recovery maps to obtain approximations for $\Lambda_{t,s}$, i.e. in Eq.~\eqref{intermediate} we change $\Lambda^{-1}_{s,0}$ by ${\Lambda^{\sigma}_{P}}_{0,s}$:
\begin{equation}
    \Phi_{t,s,0}=  \Lambda_{t,0}{\Lambda^{\sigma}_{P}}_{0,s}.
\end{equation}
Clearly this map is CP, as it is formed by a composition of CP maps. As such, if now we designate the total approximated map by 
\begin{equation}
\label{approx}
    \Lambda_{t,0}^\textrm{approx} = \Phi_{t,s,0} \Lambda_{s,0},
\end{equation}
then it will be also, by construction, a CP map.  However, as $\Phi_{t,s,0}$ may depend on the time interval $[0,s]$, the map $\Lambda_{t,0}^\textrm{approx}$ is not necessarily CP-divisible, i.e., it is not necessarily Markovian. Note that $\Lambda_{t,0}^\textrm{approx} = \Lambda_{t,s}\Lambda_{s,0}{\Lambda^{\sigma}_{P}}_{0,s} \Lambda_{s,0}$, so the non-CP map $\Lambda_{t,s}$ is still contained in $\Lambda_{t,0}^\textrm{approx}$ and this is the reason why the approximated map is not CP-divisible. Below, we analyse how well $\Lambda_{t,0}^\textrm{approx}$ approximates $\Lambda_{t,0}$, and compare the non-Markovianity of both channels.

\subsection{Non-Markovian Dephasing models}

To analyse the impact of using the Petz recovery map to obtain an approximated channel, we will exploit two dephasing, Eq.~\eqref{ch:deph}, non-Markovian models. The reference state, from here on, will thus be set to the maximally mixed one. 

\begin{figure}[H]
    \centering
    \includegraphics[width=0.9\linewidth]{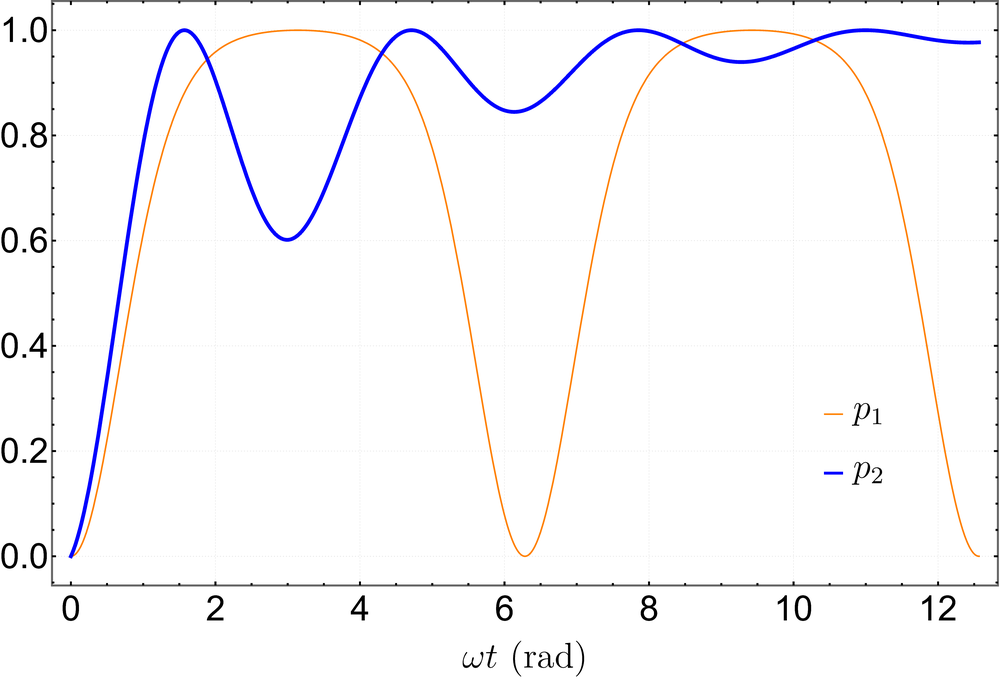}
    \caption{Probabilities  $p_{1}$ (case 1, orange line ) and $p_{2}$ (case 2, blue thick line) versus $\omega t$.}
    \label{probabilities}
\end{figure}

For the first model, named here Case 1, we construct a non-Markovian evolution by setting a time-dependent error probability as:
\begin{equation}
    \label{probability1}
    p_{1}(t) = \alpha (1 - e^{-2(1 - \cos{\omega t})}) ,
\end{equation}
where $\alpha = e^{4}/(e^{4} - 1)$, and $\omega$ is a system natural frequency (see Fig.~\ref{probabilities}). This is a periodic function with period $2\pi$, fully characterized in terms of the Lindblad local generator with negative rates \cite{Chrus1}, an explicit derivation is given in Appendix \ref{ap:generator}. A state $\rho$ undergoing this evolution reaches the maximally dephased state when $\omega t = (2n +1) \pi$, with $n \in \Nl$.

For the second model, named Case 2, the error probability will be defined as
\begin{equation}
    \label{probability2}
    p_{2}(t) = 1 - e^{- 0.3 \omega t}\cos^{2}{ \omega t}.
\end{equation}
In this case, a system undergoing this evolution reaches the maximally dephased state, goes back to an intermediate state between the maximally dephased and the initial one, and repeat the process, with the intermediate state even being closer to the maximally dephased than before. When $\omega t $ goes to infinity, the system reaches its maximally dephased state. Both probabilities $p_1$ and $p_2$ are displaced in Fig.~\ref{probabilities}.

We start our analysis by observing the so-called  ``information backflow''   that can appear in non-Markovian (NM) dynamics~\cite{Breuer2,RHP}. Concretely, take the trace distance $D(\rho_1,\rho_2)= 1/2||\rho_1- \rho_2||_1$ between two states as a quantifier of their distinguishability. For a NM dynamics it can be that $D(\Lambda_{t,0}(\rho_1), \Lambda_{t,0}(\rho_2))\le D(\Lambda_{t^\prime,0}(\rho_1), \Lambda_{t^\prime,0}(\rho_2))$, for $t^\prime \ge t$. This should be contrasted with the Markovian case, where $D(\Lambda_{t,0}(\rho_1), \Lambda_{t,0}(\rho_2))\ge D(\Lambda_{t^\prime,0}(\rho_1), \Lambda_{t^\prime,0}(\rho_2))$ for all $t^\prime \ge t$.

\begin{figure}[h]
    \centering
    \includegraphics[width=0.9\linewidth]{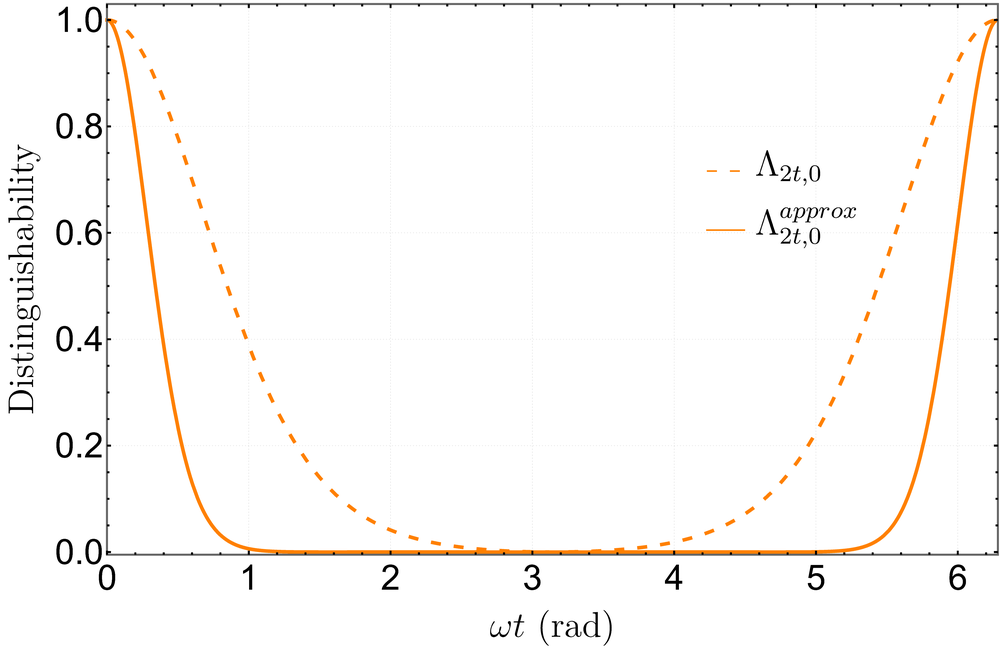}\vspace{0.5cm}
    \includegraphics[width=0.9\linewidth]{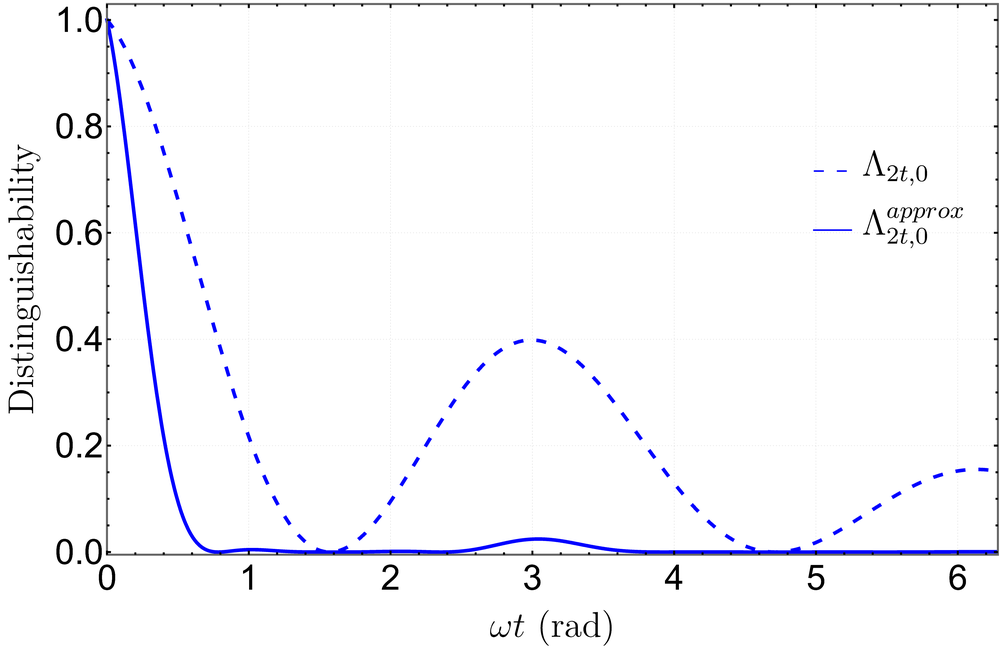}
    \caption{Distinguishability between two states $D(\rho_{1}(t),\rho_{2}(t))$ versus time, with initial states $\rho_{1} = |+\rangle\langle+|$ and $\rho_{2} = |-\rangle\langle-|$, for case 1 (top) and case 2 (bottom), where $\rho_i(t)=\Lambda_{2t,0}(\rho_i)$ (dashed line) and $\rho_i(t)=\Lambda^{approx}_{2t,0}(\rho_i)$ (solid line) with $i=1,2$. }
    \label{disting}
\end{figure}

In Fig.~\ref{disting} we compare, for both dephasing models, the  information backflow of the original map with the corresponding approximated versions. In all the cases, as initial states we took $\rho_1=\proj{+}$ and $\rho_2=\proj{-}$, as these states are orthogonal and consequently optimal states to detect information backflow.
For simplicity, we have fixed the final time as twice the intermediate time in Eq.~(\ref{approx}), i.e., $s\mapsto t$ and $t\mapsto 2t$. From Fig.~\ref{disting}, it is clear that the approximated dynamics are also non-Markovian, as both present information backflow. It is also evident that the non-Markovianity strength in both approximated cases is decreased when compared with the original maps. In case 1 the recovery of distinguishability is almost completely destroyed. On the other hand, in case 2, the complete backflow is still present in the approximated case. It is only a matter of waiting a longer time, but a similar backflow is present in the approximate and original dynamics.

In order to have a state-independent characterization, now we directly compare the original, $\Lambda_{2t,0}$, and approximated channels, $\Lambda^{approx}_{2t,0}$, for both models. 

\begin{figure}[h]
    \centering
    \includegraphics[width=0.9\linewidth]{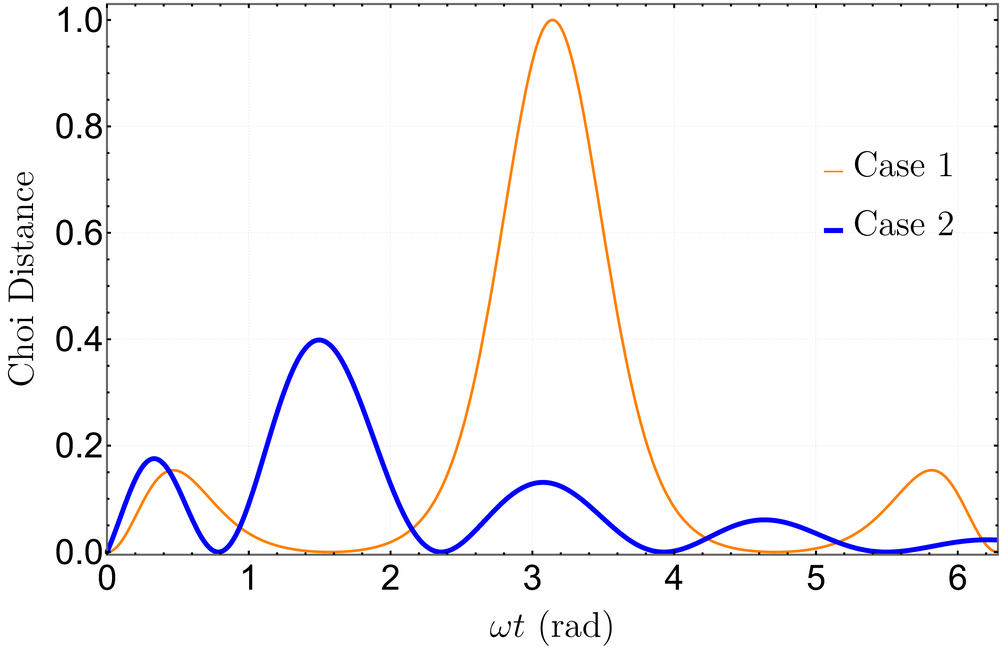}
    \caption{Normalized Choi distance $||J(\Lambda^{approx}_{2t,0}) - J(\Lambda_{2t,0})||_{1}$ between the original $\Lambda_{2t,0}$ and approximated channels $\Lambda^{approx}_{2t,0}$  versus time for case 1 (orange line) and case 2 (blue thick line).}
    \label{distances}
\end{figure}

To quantify the distance between the maps we evaluated the trace distance between their Choi matrices 
\begin{equation}
\label{choi}
     ||J(\Lambda^{approx}_{2t,0}) - J(\Lambda_{2t,0})||_{1},
\end{equation}
where $J(\Lambda)=(\mathds{1}\otimes \Lambda)(|\Omega\rangle\langle\Omega|)$ is the Choi Matrix of $\Lambda$ with $|\Omega\rangle = \frac{1}{\sqrt{2}} \sum_{i=0}^{1} |i\rangle \otimes |i\rangle$  a maximally entangled state. The norm of the Choi matrix, also known as the dynamical matrix, is widely used to quantify the degree of non-complete positiveness of a map \cite{Choi,Karol}. The results are shown in Fig.~\ref{distances}.

Combining  the results shown in Fig.~\ref{disting} and Fig.~\ref{distances}, one interesting observation emerges. 
 For a fixed value for the distances between the original and approximated maps, the recovery of the information backflow can be very different. For example, in Fig.~\ref{distances}, the distances between the dynamics at $\omega t=5.9$ rad (case 1) and in time $\omega t=3.1$ rad (case 2) are almost the same; however the information backflow displayed by the approximated channel in case 1 is closer to the original backflow than what is recovered in case 2, as can be seen in Fig.~\ref{disting}. In  experimental implementations such a result implies that approximations within a fixed distance from the theoretical dynamics may lead to very different behavior of non-Markovian features. In another way, the information backflow is not a robust non-Markovian feature.
 
\section{Conclusion}
\label{conslusion}

In this work we have explored the use of the Petz recovery map in a general context. We established the best Petz map for paradigmatic one-qubit channels (unital and non-unital) optimizing the reference states. When using as input the full set of single qubit states, we showed that the optimal Petz recovery map is worst than simply return the output state, i.e. to use  the identity channel as ``recovery channel''. We also showed that among the strategies analyzed, the dephasing channel is the decoherence channel that can be better recovered in terms of the average fidelity between the initial and the recovered state.

The Petz recovery map has been also explored in the context of non-Markovian evolution. We approximated the inverse of the map $\Lambda_{s,0}$ by its optimal Petz recovery map, and showed that the approximated evolution still presents backflow of information, but in a attenuated way. We also analyzed how the distance between the actual dynamics and the approximated one impact on the backflow of information. We observed that equally good approximations, might lead to drastically different behaviours for the backflow of information. This can be specially interesting for experimental implementations of non-Markovian evolutions, and for the use of non-Markovianity in the context of  hiding and retrieving information, as in a quantum vault \cite{qv}.

\begin{acknowledgments}
We gladly acknowledge fruitful discussions with Gabriel Landi, and thank Mark M. Wilde and Carlos Pineda for a careful reading of an earlier version of this manuscript. 
This work is supported by the Brazilian funding agencies CNPq and CAPES, and it is part of the Brazilian National Institute for Quantum Information.
\end{acknowledgments}

\appendix

\section{Optimal Petz's Recovery Map: dephasing channel}
\label{ap:analytical}

In this appendix we analytically show that the optimal Petz recovery map for the dephasing noise is obtained with an arbitrary value of the parameter $q$ in the reference state \eqref{eq:sigma}.


The first thing to notice is that $\sigma$ is diagonal in the computational basis. As such, it does not suffer the influence of the dephasing channel:
\begin{equation}
    \Lambda_{Deph}(\sigma)=\sigma,\; \forall 0\le q \le 1.
\end{equation}
Also due to this diagonal property, it is simple to obtain the reference state's square roots:
\begin{equation}
    \sigma^{\pm \frac{1}{2}}=(1-p)^{\pm \frac{1}{2}}\proj{0}+p^{\pm \frac{1}{2}}\proj{1}\;.
\end{equation}

Now, let $\rho=[\rho_{ij}]$, with $i,j\in\{0,1\}$, be a generic single qubit state. Using the results above, and remembering that the dual channel of the dephasing map is the dephasing map itself, it is now simple to show that the Petz recovery map~\eqref{Petz} for the present case is such that:

\begin{equation}
   \Lambda_{P}^\sigma(\rho) 
    = \Lambda_{Deph}(\rho). 
\end{equation}

From the equation above, we then conclude that the Petz recovery map for the dephasing channel, with reference state $\sigma$ as in \eqref{eq:sigma}, is independent of the parameter $q$. In this way,  any choice of $q$ will lead to same recovered state.   
\\
\section{Time-local generator}
\label{ap:generator}
A random unitary dynamical map 
\begin{equation}
    \Lambda_t(\rho) = \sum_{k=0}^{3} p_k(t)\sigma_k \rho \sigma_k,
\end{equation}
can be fully characterized in terms of a local generator given by 
\begin{equation}
\label{eq:generator}
\mathcal{L}(\rho) = \sum_{k=1}^{3} \gamma_{k}(t)(\sigma_k\rho\sigma_k - \rho),    
\end{equation}
where $\gamma(t)$ are the decoherence rates and $\sigma_{1},\sigma_2, \sigma_3$ are the Pauli matrices. Using the framework developed in \cite{Chrus1}, a random unitary dynamics is Markovian iff 

\begin{equation}
    \gamma_{1}(t)\geq 0, \text{ } \gamma_{2}(t)\geq 0, \text{ } \gamma_{3}(t)\geq 0, \text{for all }t\geq 0. 
\end{equation}

In order to obtain a non-Markovian evolution, we choose decay rates that can assume negative values, $\gamma(t_i)<0$ for some $t_i$. If the dynamics has only one decoherence channel, i.e only one decoherence rate is non vanishing $\gamma_k$, the time dependent probabilities can be obtained by 
\begin{equation}
    \label{eq:prob_gamma}
    p_{k}(t) = \frac{1}{2}[1 - e^{-2\Gamma_k(t)} ],  
\end{equation}
where $p_0(t) = 1 - p_{k}(t)$ and 
\begin{equation}
    \Gamma_k (t) = \int_{0}^{t} \gamma_k(\tau)d\tau. 
\end{equation}
For the first dephasing model defined in section \ref{applications}, we chose $\gamma(t) = \sin(t)$. From equations above we easily obtain 
\begin{equation}
    p_t = \alpha (1 - e^{-2(1-\cos(t))}), 
\end{equation}
where $\alpha = e^4/(e^4 - 1)$. 

For the second model, we take an oscillatory function $\gamma(t) = \frac{\cos(t)(-0.3 \cos(t) -2 \sin(t))}{e^{0.3 t} - 2\cos(t)^2}$, which gives us  
\begin{equation}
    p_{t} = 1 - (e^{-0.3 t}\cos(t)^2). 
\end{equation}

\bibliography{references}

\end{document}